\documentclass[pra,twocolumn,showpacs,preprintnumbers,amsmath,amssymb]{revtex4}

\usepackage[dvips]{graphics}
\usepackage{amsmath, amsthm, amssymb}
\usepackage{graphicx}
\usepackage{dcolumn}
\usepackage{bm}

\newcommand{\bra}[1]{\langle #1|}
\newcommand{\ket}[1]{|#1\rangle}
\newcommand{\braket}[2]{\langle #1|#2\rangle}

\newtheorem{lemma}{Lemma}

\newcommand{\ie}{{\it{i.e.}}: }

\newcommand{\pla}[3]{{Phys. Lett A} {#1}, {#2} (#3)}

\newcommand{\jpa}[3]{{J. Phys. A: Math. Gen.} {\bf{#1}}, #2 (#3)}

\newcommand{\quantph}[2]{{Preprint} quant-ph/#1 (#2)}

\newcommand{\tr}{\mathrm{Tr}}

\begin{document}

\title{Connecting the generalized robustness and the geometric
measure of entanglement}

\author{Daniel Cavalcanti}

\affiliation{Departamento de F\'{\i}sica, Caixa Postal 702,
Universidade Federal de Minas Gerais, 30123-970, Belo Horizonte, MG,
Brazil} \email{dcs@fisica.ufmg.br}
\begin{abstract}
The main goal of this paper is to provide a connection between the
generalized robustness of entanglement ($R_g$) and the geometric
measure of entanglement ($E_{GME}$). First, we show that the
generalized robustness is always higher than or equal to the
geometric measure. Then we find a tighter lower bound to $R_g(\rho)$
based only on the purity of $\rho$ and its maximal overlap to a
separable state. As we will see it is also possible to express this
lower bound in terms of $E_{GME}$.
\end{abstract}
\pacs{03.67.-a, 03.65.Ud, 03.67.Mn}

\maketitle

Since it was first noted \cite{Sch,EPR} the issue of quantum
correlations has been largely studied and debated. However, it was
not until entanglement was recognized as a physical resource that
this theme got a solid status. From this point of view, entanglement
was shown to allow several tasks such as quantum cryptography
\cite{Ekert}, teleportation \cite{telep},  and quantum algorithms
\cite{Nielsen}. On the other hand, entanglement has also given us
new insights for understanding important physical phenomena
including superconductivity \cite{high-tc}, super-radiance
\cite{Dicke}, quantum phase transitions
\cite{QPT1,QPT2,QPT3,QPT4,QPT5}, and the appearance of classicality
\cite{Zurek}.

One of the greatest challenges concerning entanglement is how to
properly quantify this resource. Although this problem is well
understood for bipartite pure states, in a more complex scenario
(multipartite systems or mixed states) a complete theory on the
quantification of entanglement is still lacking.

Among the difficulties of dealing with multipartite entanglement is
the fact that systems composed by various parts can exhibit many
kinds of entanglement. This is because one may be interested in the
entanglement according to a specific partition of the whole system.
So a state can present some entanglement in relation to a given
partition, while it can be separable according to another one.

In the last years many candidates of entanglement quantifiers were
proposed. Generically speaking, the ways of quantifying entanglement
can be divided into two classes: quantifiers with a geometrical
interpretation, and those with an operational meaning. In the first
class we can cite the relative entropy of entanglement
\cite{vedral1,vedral2}, the geometric measure of entanglement
\cite{GME,wei1}, the negativity \cite{negat1,negat2,negat3}, and the
robustness of entanglement \cite{robust,Ste1,Ste2}. The entanglement
cost \cite{cost,cost2}, the distillable entanglement
\cite{cost,dist}, and the singlet fraction \cite{singfrac} are
examples of operational measures.

The purpose of this letter is to point out a connection between two
well discussed entanglement quantifiers, the generalized robustness
($R_g$) \cite{Ste1,Ste2} and the geometric measure of entanglement
($E_{GME}$) \cite{GME,wei1}. That these quantifiers are related is
not obvious {\it{a priori}}, for these functions rely on distinct
geometrical interpretations. While $E_{GME}$ measures the minimum
angle between an entangled state and a separable one, $R_g$ can be
treated as a kind of ``distance" between an entangled state and the
set of unentangled states. Furthermore both quantifiers are able to
deal with the various types of entanglement that a multipartite
system can present.

Let us first present the language we shall adopt to talk about
multipartite entanglement. Suppose a state $\rho$ can be written as
a convex combination of states which are product of $k$ tensor
factors. The state $\rho$ is then said to be a $k$-separable state.
One should note that in a system of $n$ parts, $n$-separability is
separability itself and that every state is trivially $1$-separable.
The set of $k$-separable states will be denoted by $S_k$. It is
clear that $S_n \subset S_{n-1} \subset...\subset S_{1} = D$, where
$D$ denotes the set of density operators.

We are now able to understand why $R_g$ and $E_{GME}$ can
distinguish the types of entanglement a system contains. The
geometric measure is a pure-state entanglement quantifier given by:
\begin{equation}
E_{GME}^k(\psi)=1-\Lambda_{k}^2(\psi),
\end{equation}
where
\begin{equation}
\Lambda_{k}^2=\max_{\phi \in S_{k}}|\braket{\phi}{\psi}|^{2}.
\end{equation}
Thus $E_{GME}^k(\psi)$ measures the sine squared of the minimum
angle between $\ket{\psi}$ and a $k$-separable state\footnote{The
extension of $E_{GME}^k(\psi)$ to mixed states is made through the
convex-roof construction, which makes this quantifier hard to be
computed in general. Furthermore this construction makes the
geometrical interpretation of $E_{GME}^k$ not so clear. Thus we will
focus on pure states unless otherwise specified.}. It is known that
this quantity is an entanglement monotone \cite{monot}, \ie it is a
non-increasing function under LOCC.

The relation between $E_{GME}^k$ and the notion of $k$-entanglement
witnesses \cite{HHH} (observables with positive mean value to all
$k$-separable states, but negative to some $\rho \notin S_k$) (see
Ref. \cite{wei1}) has also been determined. This results from the
fact that one can always construct a $k$-entanglement witnesses
$W^k$ for a pure state $\ket{\psi}$ of the type
\begin{equation}\label{Wit}
W^k=\lambda^2-\ket{\psi}\bra{\psi}.
\end{equation}
As this operator must have a positive mean value for every
$k$-separable state, the relation
\begin{equation}
\lambda^2\geq\max_{\ket{\phi} \in
S_k}\|\braket{\phi}{\psi}\|^2=\Lambda_{k}^2
\end{equation}
must hold. Thus the optimal entanglement witness of the form
(\ref{Wit}) is reached when $\lambda=\Lambda_{k}^2$, and we can
write
\begin{equation}\label{test}
W_{opt}^k=\Lambda_{k}^2-\ket{\psi}\bra{\psi}.
\end{equation}
Here optimality is defined in the sense of getting the highest value
to $|\bra{\psi}W^k\ket{\psi}|$.

\begin{figure}\centering
 {\includegraphics[scale=0.40]{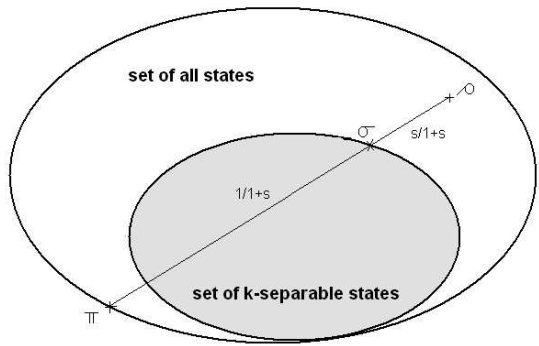}}
\caption{Geometrical interpretation of $R^{k}_g$. The straight line
represents the convex combination $\frac{\rho+s\pi}{1+s}$. We see
that for a given state $\pi$ and a value of $s$ this combination
becomes $k$-separable. $R_g^{k}(\rho)$ is defined as the minimum
$s$, considering all possible states $\pi$.}
\label{GenRob}\end{figure}

In a different fashion, the robustness of entanglement of a state
$\rho$ quantifies how robust the entanglement of $\rho$ is  under
presence of noise. Thus the robustness of $\rho$ in relation to the
state $\pi$, $R(\rho\|\pi)$, is the minimum $s$ such that the state
\begin{equation}
\sigma=\frac{\rho+s\pi}{1+s} \label{rob}
\end{equation}
is $k$-separable. We will be interested in an extension of the
relative robustness, namely the {\emph{generalized robustness}}.
This entanglement quantifier is obtained by the minimization of the
relative robustness over all states $\pi$ \cite{Ste1}. Recently, an
interesting operational interpretation to $R_{g}^k$ was given in
terms of the percentual increase a state can provide to
teleportation processes \cite{Bra1}. The generalized robustness can
also be viewed as a ``distance'' of $\rho$ to the set $S_k$ in the
space of states (see figure \ref{GenRob}) \cite{GPT}, and thus allow
both a geometrical and an operational interpretation. Moreover
$R_{g}^k$ was used to investigate the shape of entangled states sets
\cite{GPT} and was shown to exhibit a kind of {\emph{polygamy of
entanglement}} \cite{CBT05}.

As well as the geometric measure, $R_{g}^k$ is intimately connected
to the notion of entanglement witnesses. In fact, $R^{k}_g$ can be
expressed as
\begin{equation}\label{Ew}
R_{g}^{k}(\rho)=\max \{0, -\min_{W^{k} \in
\mathcal{M}}\tr(W^{k}\rho)\},
\end{equation}
where $\mathcal{M}$ is the set of operators $M \leq I$ and $W^k$ is
a $k$-entanglement witness \cite{Bra}.

As the witness (\ref{test}) obviously  satisfies the condition
$W^k\leq I$ we can attest the following:
\begin{equation}\label{ineq}
R_g^{k}(\psi)\geq E_{GME}^{k}(\psi).
\end{equation}
Some points concerning the inequality (\ref{ineq}) should be
stressed at this stage. First, it is a relation valid to all kinds
of multipartite entanglement. Moreover this relation will be strict
whenever the witness (\ref{test}) is a $k$-entanglement witness
which solves the minimization problem in (\ref{Ew}). Finally, one
could argue that relation \eqref{ineq} may be, in fact, a
consequence of standard results from matrix analysis relating
different distance measures between operators (as commented, both
$R_g^{k}$ and $E_{GME}^{k}$ are related to such distances). However,
it must be clear that $R_g^{k}(\psi)$ is not simply the distance
between $\psi$ and its closest state $\sigma \in S_k$, but one
should keep in mind that this distance is taken with relation to the
state $\pi$ as a reference \footnote{Besides that there is a
minimization among all possible states $\pi$.} (recall figure 1).
This makes the closest $k$-separable state usually different for
$R_g^{k}$ and $E_{GME}^{k}$.

In fact, it is possible to give a tighter relation between $R_{g}^k$
and $E_{GME}^k$. Recall the lemma 1 shown in ref. \cite{CBT05}. We
now give a clearer proof of it, and interpret it as a lower bound to
$R_{g}^k$.
\begin{lemma}
For every state $\rho \in D$,
\begin{equation}\label{tighter}
R_{g}^{k}(\rho) \geq \frac{\tr(\rho^2)}{\max_{\sigma \in
S_{k}}\tr(\rho \sigma)}-1.
\end{equation}
\end{lemma}
\textbf{Proof} First of all let us show that $\max_{\sigma \in
S_{k}}\tr(\rho \sigma)$ is equal to the minimum value of $\lambda$
($\lambda_{min}$) such that $W=\lambda I-\rho$ is a $k$-entanglement
witness. As $\tr(W \sigma)\geq 0 \hspace{0.04cm}\forall
\hspace{0.04cm} \sigma \hspace{0.02cm} \in\hspace{0.02cm} S_k$,
\begin{equation}
\tr[(\lambda I-\rho)\sigma]=\lambda-\tr(\rho\sigma)\geq0.
\end{equation}
It is thus straightforward to see that $\lambda_{min}=\max_{\sigma
\in S_{k}}\tr(\rho \sigma)$.

Note that
\begin{equation}
W'=\frac{W}{\lambda_{min}}=I-\frac{\rho}{\lambda_{min}}<I.
\end{equation}
So it is possible to see that $R_{g}^k(\rho)\geq -\tr(W' \rho)$,
from which follows the required result. $\square$

The lower bound to $R_g^k$ expressed by (\ref{tighter}) can be
easily interpreted:  $\tr{\rho^2}$ measures the purity of $\rho$,
and $\tr(\rho \sigma)$ is the Hilbert-Schmidt scalar product between
$\rho$ and $\sigma$. It is expected that the more mixed $\rho$ is,
the lower the value of $\tr{\rho^2}$ gets, and the state becomes
less entangled. Similarly, the larger $\max_{\sigma \in
S_{k}}\tr(\rho \sigma)$ is, closer to the set $S_k$ $\rho$ gets, and
the system will show less entanglement.

But now we note that in the special case of pure states the
relations $\tr(\rho^2)=1$ and $\max_{\sigma \in S_{k}(H)}\tr(\rho
\sigma)=\Lambda_{k}^2(\rho)$ hold and therefore we have the general
relation
\begin{equation}\label{RgLam}
R_{g}^{k}(\psi) \geq \frac{1}{\Lambda_{k}^2(\psi)}-1.
\end{equation}
and we can see the relation we are looking for:
\begin{equation}
R_{g}^{k}(\psi) \geq \frac{E_{GME}^k}{1-E_{GME}^k}.
\end{equation}

It is interesting that two entanglement monotones with different
geometric interpretation are actually related, and furthermore this
relation allows an analytic lower bound to the generalized
robustness for all states whenever $\Lambda_{k}^2(\rho)$ can be
analytically computed. This is the case, for example, of completely
symmetric states, Werner states, and the isotropic states
\cite{wei1,wei2}.

\begin{figure}\centering
 \rotatebox{270}{\includegraphics[scale=0.45]{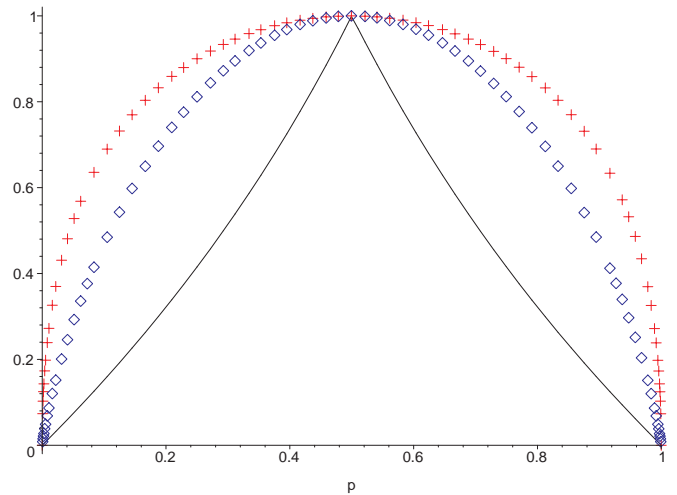}}
\caption{(Color online) \textbf{Red crosses:} logarithmic
generalized robustness of entanglement. \textbf{Blue diamonds:}
relative entropy of entanglement. \textbf{Black line:} lower bound
given in Eq. \eqref{LGR}.} \label{comp}\end{figure}

We can furthermore see from (\ref{RgLam}) that
\begin{equation}\label{LGR}
\log_2(1+R^{k}_{g})\geq -2\log_2\Lambda_{k}.
\end{equation}
The left side of this expression is the logarithmic robustness of
entanglement ($LR_g^k$), another entanglement quantifier with
interesting features \cite{Bra}. Curiously, this is exactly the same
lower bound expressed to the relative entropy of entanglement
($E_R^k$) in \cite{wei2}. Numerical and analytical results (see, for
example, Figure \ref{comp} and Table \ref{table1}) suggest that
$LR_g^k\geq E_R^k$, in general, but at the moment this is just a
conjecture.

For bipartite pure states all the quantities considered so far can
be analytically computed. While the relative entropy of entanglement
equals the entropy of entanglement (given by the von Neumann entropy
of the reduced state) \cite{vedral2}, the generalized robustness is
given by
\begin{equation}
R_{g}^k(\psi)=(\sum_i c_i)^2 -1,
\end{equation}
being $\{c_i\}$ the spectrum of Schmidt of $\ket{\psi}$ \cite{Ste1}.
In this context it can be noted that $\Lambda_k$ is given by the
modulus of the highest Schmidt coefficient of $\ket{\psi}$
\cite{wei1}. To visualize and compare these entanglement measures we
calculate the relative entropy of entanglement, the logarithmic
generalized robustness, and the lower bound expressed in \eqref{LGR}
for the state
\begin{equation}
\ket{\psi(p)}=\sqrt{p}\ket{00}+\sqrt{1-p}\ket{11}.
\end{equation}
The plots are available in figure \ref{comp}.

As the presented relations between $R_g^k$ and $E_{GME}^k$ are also
valid to multipartite entanglement it would be useful to illustrate
the results in this context as well. We choose to study some
completely symmetric states for this aim. These states are referred
to as Dicke states or, some times, as generalized $W$ states and
appear naturally as eigenstates of various models such as the
$\eta$-pairing model \cite{high-tc} and the Dicke model
\cite{Dicke}. Following ref. \cite{wei1} we will label these states
according to the number of 0's, as follows:
\begin{equation}\label{Ws}
\ket{S(n,k)}=\sqrt{\frac{k!(n-k)!}{n!}}S\ket{\underbrace{000..0}_{k}\underbrace{11..1}_{n-k}},
\end{equation}
where $S$ is the total symmetrization operator. Wei and Goldbart
showed an analytical expression to $E_{GME}^n(\ket{S(n,k)})$ (\ie
the geometric measure of $\ket{S(n,k)}$ with relation to the
completely separable states) \cite{wei1}. Additionally, in this case it was shown
that the relative entropy of entanglement is exactly equal the lower
bound given in Eq. \eqref{LGR} \cite{wei2}, and moreover it equals the logarithmic robustness of entanglement \cite{Damian}. 
So, for the states \eqref{Ws}, it is possible to compare analytically these entanglement quantifiers. Some examples are shown in Table \ref{table1}.
\begin{table}
  \centering
  \begin{tabular}{|c|c|c|c|c|}
    \hline
      & $\ket{S(2,1)}$ & $\ket{S(3,2)}$ & $\ket{S(4,3)}$ & $\ket{S(4,2)}$ \\
    \hline
    $E_{GME}^n$ & 0.5 & 0.55 & 0.58 & 0.625 \\
    $R_g^n$ & 1 & 1.25 & 1.36 & 1.65 \\
    \hline
  \end{tabular}
  \caption{A comparison among multipartite entanglement of some states \eqref{Ws},
  given by  geometric measure of entanglement ($E_{GME}^n$) - see Ref. \cite{wei1} -
and the robustness of entanglement ($R_g^n$) - see Ref. \cite{Damian}.}\label{table1}
\end{table}

In brief, we have shown some relations between the geometric measure
of entanglement and the generalized robustness of entanglement. We
reached a lower bound to $R_{g}^k$ with nice interpretations and
wrote it in terms of $E_{GME}^k$. These relations also allowed us to
compare two other entanglement quantifiers, the logarithmic
generalized robustness and the relative entropy of entanglement.
Examples were given to illustrate the results.

Because many entanglement quantifiers exist it is important to
understand their relation and this, we believe, should be a major
goal in the theory of entanglement. We hope that this discussion can
help in this sense.

\begin{acknowledgments}
I would like to thank Dr. Marcelo Terra Cunha and Flavia Tenuta for
useful comments on this paper. I specially thank Dr. Miguel G. Cruz
for helpful criticisms, Fernando Brand\~ao for help with some
numerical results, and Damian Markham for bringing ref. \cite{Damian} to my attention. Financial support from CNPq is also acknowledged.
\end{acknowledgments}

\end{document}